\documentclass{DISproc}

\usepackage{subfig}
\usepackage{units}

\begin{document}
\title{A Quick Tour of Ultra-Relativistic Heavy-Ion Physics at the LHC}

\author{{\slshape Klaus Reygers$^1$, for the ALICE collaboration}\\[1ex]
$^1$Physkalisches Institut, Universit{\"a}t Heidelberg, Im Neuenheimer Feld 226, 69120 Heidelberg, Germany}

\contribID{xy}

\doi  

\maketitle

\begin{abstract}
 A brief summary of results on Pb+Pb collisions from ALICE, ATLAS, and CMS is presented covering global event properties, anisotropic flow, jet quenching, and quarkonia. 
\end{abstract}

\section{Introduction}
The objective in ultra-relativistic heavy-ion physics is to study the strong interaction in the limit of high temperatures and densities. Quantum Chromodynamics, the theory of the strong interaction, is experimentally well-tested in the limit of weak coupling and low parton densities (perturbative QCD) and in the limit strong coupling for static systems at vanishing temperature (lattice QCD at $T \approx 0$). Examples are the description of jet spectra and the determination of hadron masses in lattice QCD, respectively. In heavy-ion physics one explores the regime of strong coupling at $T \gg 0$. 

A prediction for this regime from first QCD principles is a transition from confined to deconfined quarks and gluons, i.e., to a quark-gluon plasma (QGP), at a temperature of $T_c \approx \unit[150-160]{MeV}$ or roughly $\unit[1.8\cdot10^{12}]{K}$ \cite{Borsanyi:2010cj,Bazavov:2011nk}. This corresponds to an energy density of $\varepsilon_c \approx \unit[0.5]{GeV/fm^3}$. This prediction from lattice QCD holds for a baryo-chemical potential $\mu_B = 0$, i.e., for systems with equal numbers of quarks and anti-quarks. These calculations indicate that the confinement/deconfinement transition at $\mu_B = 0$ is a cross-over transition. 

At larger $\mu_B$ there could be a critical endpoint in the QCD phase diagram where the cross-over transition turns into a first-order transition. The search for features like this in the QCD phase diagram motivates the RHIC beam energy scan program and future experiments at FAIR. For the matter created at full RHIC energy and at the LHC $\mu_B \approx 0$.

The LHC plays a crucial role in heavy-ion physics \cite{Muller:2012zq}: The increase in energy by a factor $\sim 14$ from $\sqrt{s_{NN}} = \unit[0.2]{TeV}$ at RHIC to $\sqrt{s_{NN}} = \unit[2.76]{TeV}$ at the LHC provides very high initial energy densities ($\varepsilon \gtrsim \unit[15]{GeV/fm^3}$) and a longer lifetime of the deconfined quark-gluon matter. Moreover, the abundant production of hard probes, like jets and heavy quarks, simplifies their use as tools to probe the medium.

The heavy-ion program at RHIC provides strong evidence for the formation of a QGP \cite{BraunMunzinger:2007zz,Tannenbaum:2012ma}. The QGP at RHIC can be characterized as an almost ideal fluid, i.e., as a strongly coupled fluid with small viscosity-to-entropy density ratio $\eta/s$. Based on anisotropic flow measurement at RHIC $\eta/s$ was found to be close to a conjectured theoretical lower bound of $(\eta/s)_\mathrm{min} = 1/4 \pi$ (in natural units) with an upper limit of about $\eta/s \lesssim 2.5/(4 \pi)$ \cite{Muller:2012zq}. Moreover, the created medium turned out to be rather opaque to quark and gluon jets, a phenomenon called ``jet quenching''. After the phase of discoveries at RHIC the objective now is to characterize the medium in a quantitative way. Furthermore, the large increase in energy at the LHC allows to test models developed based on RHIC data.

The mission of heavy-ion physics may be characterized by the following two aspects: 1)~Understand the complex phenomenology of A+A collisions and 2)~based on that, learn something about QCD thermodynamics. In more detail, one would like to find a compelling proof of deconfinement and learn something about the equation of state, the relevant number of degrees of freedom, the viscosity, the velocity of sound, etc. As jet quenching is a prominent phenomenon, the mechanism of parton energy loss is also in the focus of interest.

A standard reaction model has emerged from the study of ultra-relativistic heavy-ion collisions at the CERN SPS and at RHIC \cite{Muller:2012zq}. In the first stage of a A+A collision partons are liberated from the nuclear wave function. After a time of about \unit[1]{fm/$c$} the system has thermalized such that the concept of temperature becomes meaningful. The large pressure of the partonic matter leads to longitudinal and transverse expansion which can be described by almost ideal hydrodynamics. At the pseudo-critical temperature of $T_c \approx \unit[150-160]{MeV}$ the transition from the QGP to a hadron gas occurs. The hadron gas cools down further and at a temperature rather close to $T_c$ the relative abundances of different hadron species are fixed. This is called chemical freeze-out. After the chemical  freeze-out momentum spectra of the different particle species still change. At a temperature of about $\unit[100]{MeV}$ the densities become so low that the hadrons cease to interact. This is the kinetic freeze-out.

At RHIC, PHENIX and STAR continue to explore heavy-ion collisions with the aim to quantitatively characterize the created medium. At the LHC, ALICE, ATLAS, and CMS take part in the heavy-ion program. ALICE is the dedicated heavy-ion experiment at the LHC. It provides robust tracking over a larger $p_T$ range ($\sim \unit[0.1]{GeV} < p_T < \unit[100]{GeV}$), good primary and secondary vertex reconstruction, and excellent particle identification, especially at low $p_T$ where the bulk of the particles are produced. ATLAS and CMS are ideal for studying hard probes, e.g.,  jet production at high $p_T$. 

\section{Global Event Properties}
\begin{figure}[t]
\centering
\subfloat{\includegraphics[width=0.47\linewidth]{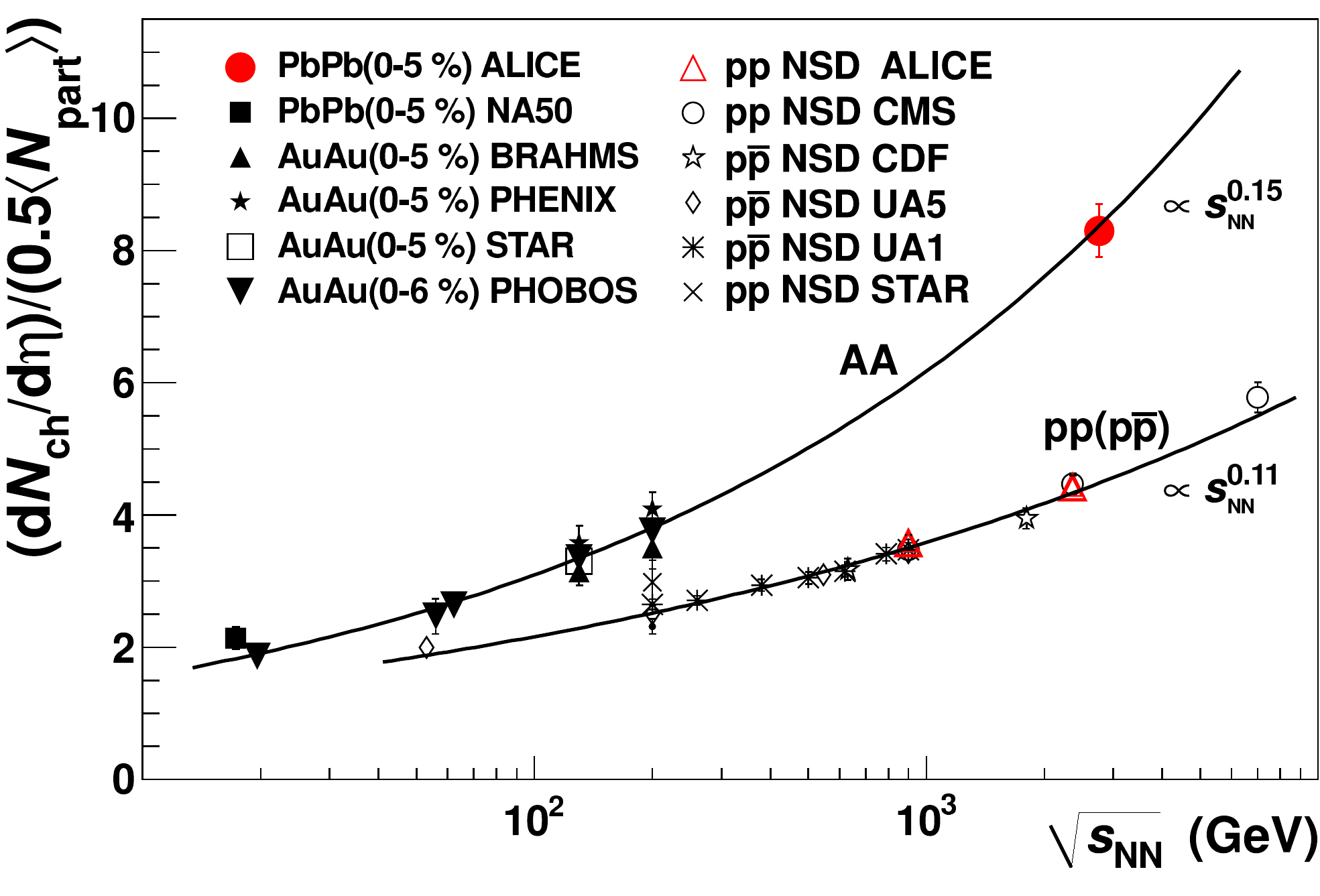}}
\hfill
\subfloat{\includegraphics[width=0.46\linewidth]{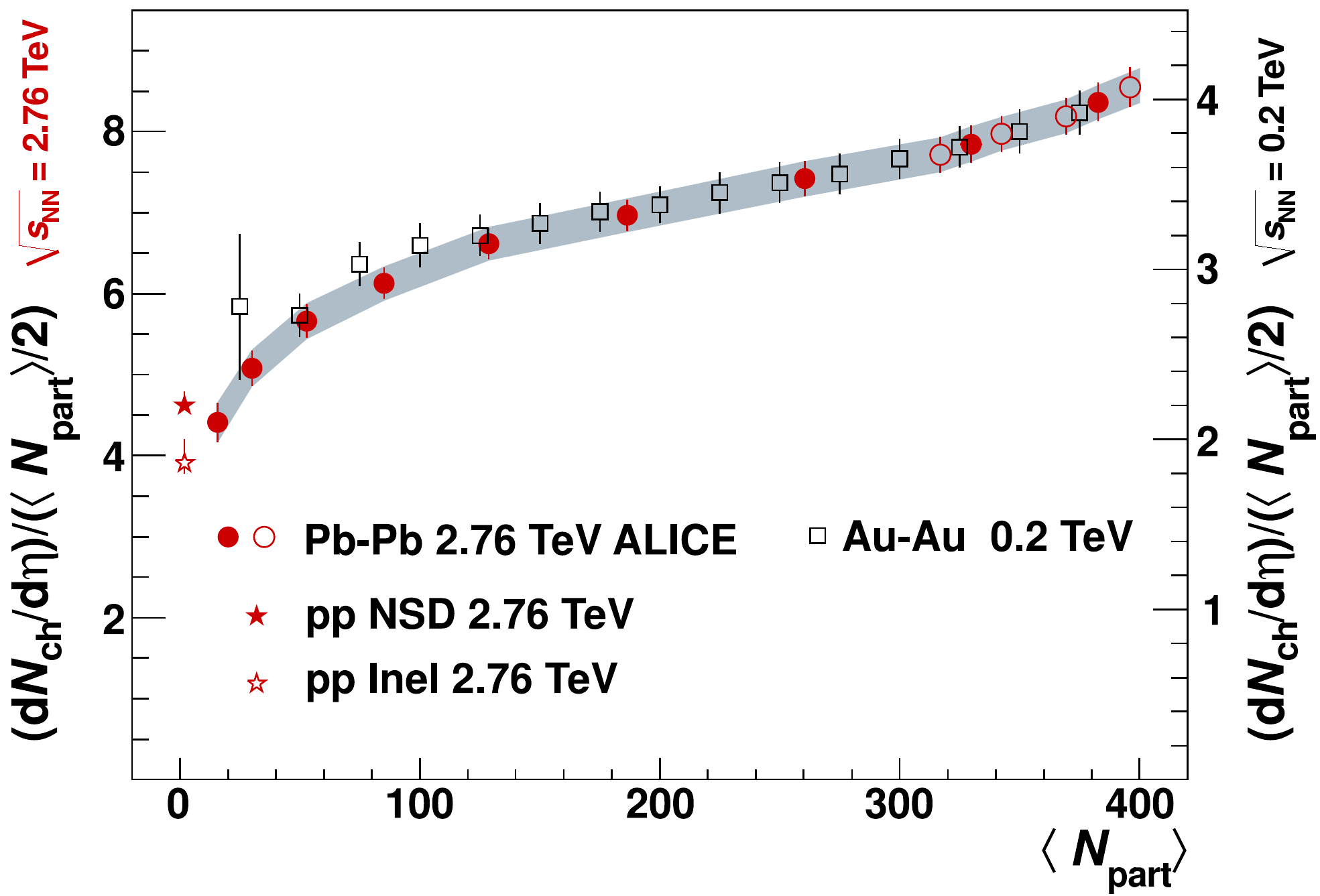}}
\caption{a) Charged-particle multiplicity normalized to $N_\mathrm{part}/2$ in pp and central A+A (Au+Au and Pb+Pb) collisions as a function $\sqrt{s_{NN}}$ \cite{Aamodt:2010pb}. The increase in central A+A collisions with $\sqrt{s_{NN}}$ is stronger than in pp collisions. b) The increase of the charged-particle multiplicity per participant pair with $N_\mathrm{part}$ at RHIC and LHC exhibits a very similar shape \cite{Aamodt:2010cz}.}
\label{fig:nch}
\end{figure}
In heavy-ion collisions, the number of produced charged particles is tightly correlated with centrality, i.e., with the impact parameter of the collision. The centrality of a collision is typically defined by measuring the charged-particle multiplicity or the transverse energy a few units of pseudo-rapidity away from mid-rapidity. In order to compare experimental results with theory calculations or to compare results from different experiments, the centrality is often expressed in terms of the number of participating nucleons ($N_\mathrm{part}$), which is calculated with a geometrical Glauber model \cite{Miller:2007ri}.

The measurement of the charged-particle multiplicity around mid-rapidity provides a first estimate of the energy density created in the central rapidity region. Fig.~\ref{fig:nch}a shows the increase of $(dN_\mathrm{ch}/d\eta)/(N_\mathrm{part}/2)$ with the center-of-mass energy (per nucleon-nucleon pair) $\sqrt{s_{NN}}$ for p+p and central A+A (Au+Au, Pb+Pb) collisions. The increase in pp and A+A collisions can be described by a power law. Interestingly, the increase with $\sqrt{s_{NN}}$ in central A+A collisions ($\sim s^{0.15}$) is stronger than in pp collisions ($\sim s^{0.11}$). With the aid of the Bjorken formula 
\begin{equation}
\varepsilon = \frac{\mathrm{d}E_T/\mathrm{d}y}{\tau_0 \pi R^2}
\approx \frac{3}{2} \langle m_T \rangle \frac{\mathrm{d}N_\mathrm{ch}/\mathrm{d}\eta}{\tau_0 \pi R^2}, 
\end{equation}
where $R \approx \unit[6.62]{fm}$ is the radius of a Pb nucleus, the initial energy density in central Pb+Pb collisions at $\sqrt{s_{NN}} = \unit[2.76]{TeV}$ can be estimated to be $\varepsilon_\mathrm{LHC} \approx \unit[15]{GeV/fm^3} \approx 3 \times \varepsilon_\mathrm{RHIC}$. This is the estimate for a thermalization time of $\tau_0 = \unit[1]{fm}/c$. The actual thermalization time is most likely smaller so that this estimate is considered conservative. Thus, one can conclude that the initial energy densities at RHIC and the LHC are well above the critical energy density of $\varepsilon_c \approx \unit[0.5]{GeV/fm^3}$ for the transition to the QGP.

The relative increase of the charged-particle multiplicity with $N_\mathrm{part}$ is rather similar at RHIC and the LHC. This is shown Fig.~\ref{fig:nch}b. This similarity has actually been demonstrated for an even wider energy interval of $20 \lesssim \sqrt{s_{NN}} \lesssim \unit[2760]{GeV}$ \cite{Alver:2010ck}. The fact that the relative increase of $dN_\mathrm{ch}/d\eta$ is rather independent of $\sqrt{s_{NN}}$ is surprising in two-component models which assume that a soft component of the multiplicity scales with $N_\mathrm{part}$ and a hard scattering component with $N_\mathrm{coll}$, the number of inelastic nucleon-nucleon collisions. In these models, the increase of the hard-scattering cross sections with $\sqrt{s_{NN}}$ needs to be balanced by shadowing of the gluon distribution function in the nucleus. Irrespective of the theoretical description, the data indicate that at a given $\sqrt{s_{NN}}$ the charged-particle multiplicity is largely determined by the geometry of the collision.

\begin{figure}[t]
\centering
\subfloat{\includegraphics[width=0.5\linewidth]{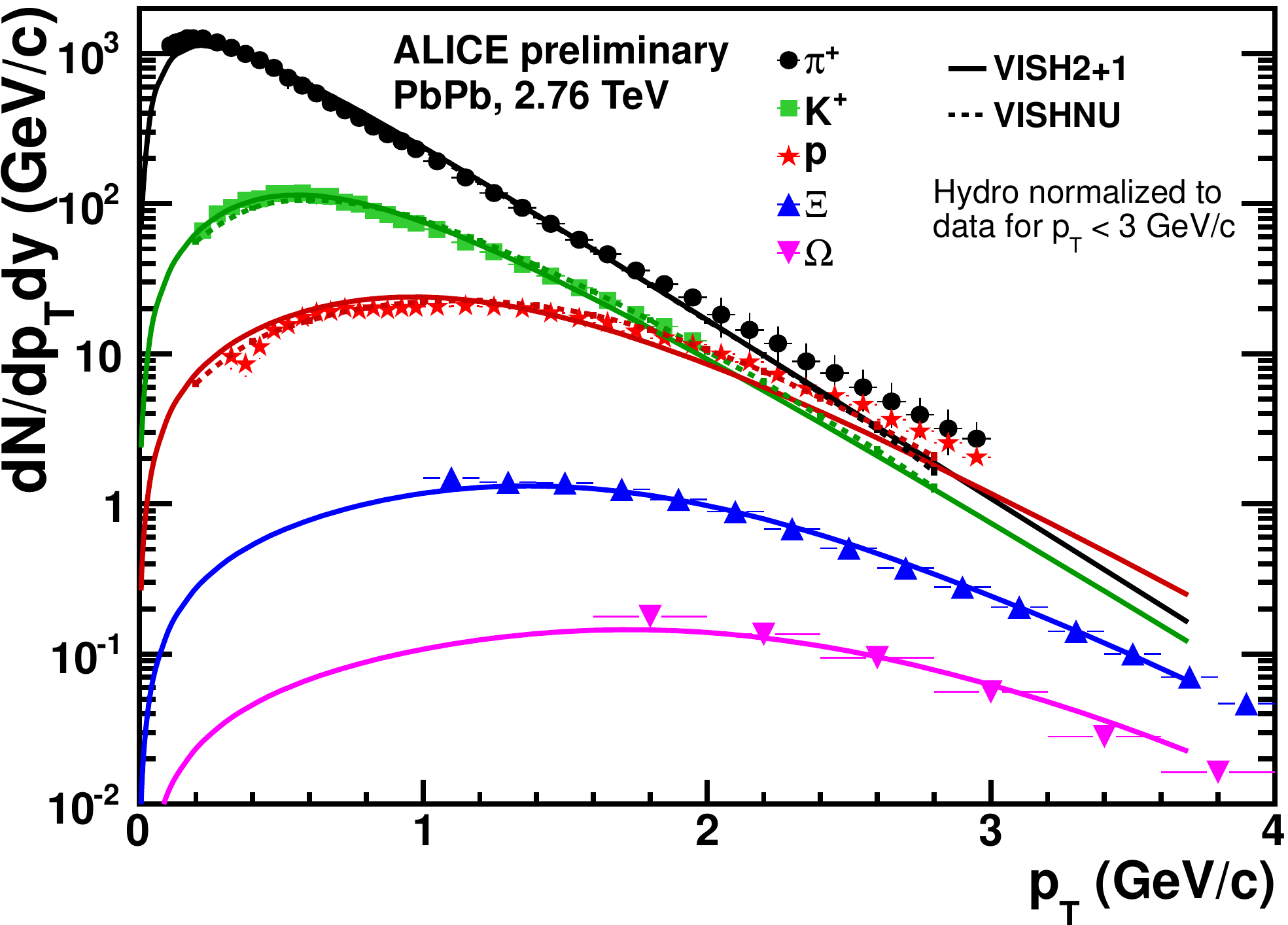}}
\subfloat{\includegraphics[width=0.5\linewidth]{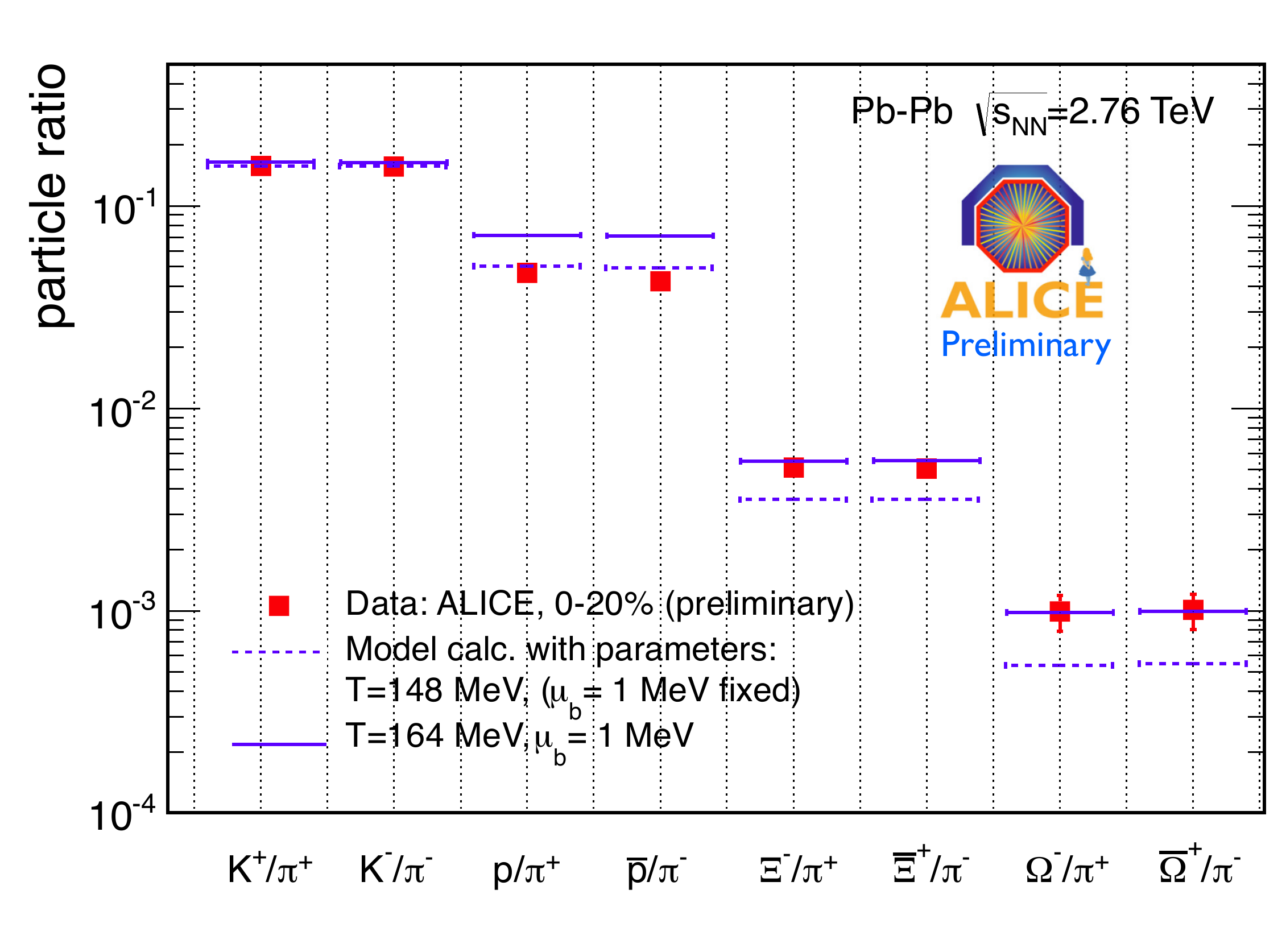}}
\caption{a) $p_T$ spectra of identified particles measured by ALICE \cite{Muller:2012zq}. The data are rather well described by hydro-models which model the expansion of the fireball. b) Ratios of the yields of different particles species measured by ALICE in comparison with predictions from a statistical model. Except for the p/$\pi$ ratios the data are well described with a chemical freeze-out temperature of \unit[164]{MeV} and a small chemical potential of $\mu_B = \unit[1]{MeV}$.}
\label{fig:pt_and_stat_mod}
\end{figure}
The shapes of transverse momentum spectra in central collisions for particles with different masses provide compelling evidence for radial flow, i.e., for a collective flow profile characterized by velocity vectors all pointing away from the center of the overlap zone with magnitudes independent of the azimuthal angle $\varphi$. Radial flow leads to a modification of the $p_T$ spectrum according to $p_T^\mathrm{w/\;flow} = p_T^\mathrm{w/o\;flow} + \beta_{T,\mathrm{flow}} \gamma_{T,\mathrm{flow}} m$ where $\beta_{T,\mathrm{flow}}$ is the radial flow velocity and $m$ the particle mass. Fig.~\ref{fig:pt_and_stat_mod} shows that the effect of radial flow is most visible for heavy particles.

The good agreement between the $p_T$ spectra of identified particles and hydrodynamical models provides the strongest evidence for radial flow, see Fig.~\ref{fig:pt_and_stat_mod}a. Essential features of complete hydro calculations are captured with so-called  blast-wave fits, which have the average transverse flow velocity $\langle \beta_{T,\mathrm{flow}} \rangle$ and the kinetic freeze-out temperature $T_{fo}$ as free parameters. The $p_T$ spectra in central Pb+Pb collisions at $\sqrt{s_{NN}} = \unit[2.76]{TeV}$ are described with $\langle \beta_{T,\mathrm{flow}} \rangle_\mathrm{LHC} \approx 0.65\,c$ and $T_{fo} \approx \unit[80 - 100]{MeV}$. The average radial flow velocity at the LHC is roughly 10\,\% larger than in central Au+Au collisions at RHIC.

Comparisons of the $p_T$-integrated yields of different particles species at RHIC and LHC with predictions from statistical models show that the chemical freeze-out temperature $T_\mathrm{ch}$ at RHIC and LHC is close to the critical temperature $T_c$ from lattice QCD. In statistical models for particle production the yield of a particle species depends on its mass and a chemical potential given by the baryon number, the strangeness, and the isospin of the particle. With two free parameters, $T_\mathrm{ch}$ and a baryo-chemical potential $\mu_B$, particle ratios are described rather well, including the enhancement for strange particles relative to pp collisions. A fit to particle ratios at the LHC yields $T_\mathrm{ch} \approx \unit[164]{MeV}$. The discrepancies for the p/$\pi$ ratios (see Fig.~\ref{fig:pt_and_stat_mod}b) remain to be understood.

\section{Anisotropic Flow}
In non-central A+A collisions the initial spatial anisotropy of the overlap zone is expected to lead to an anisotropy of produced particles in momentum space. The line connecting the two centers of the nuclei in the transverse plane defines the so-called reaction plane.  Elliptic flow results from pressure gradients being larger in the direction of the reaction plane than perpendicular to it. It was realized that fluctuations of the initial distribution of the energy density are important. The overlap zone may, e.g., look rather like a triangle in some collisions which leads to a corresponding modulation of particles in momentum space. In general, particle production in the transverse plane as a function of the azimuthal angle $\varphi$ can be described as
\begin{equation}
E \frac{{\rm d^3} N}{{\rm d^3}p} =
\frac {1}{2\pi} \frac{{\rm d^2}N}{p_{T} {\rm d}p_{T}{\rm d}y} {\left ( 1 + 2 \sum_{n=1}^{\infty} v_n \cos{[n(\varphi - \Psi_{n})]} \right )}
\end{equation}
where $v_2$ describes the elliptic flow and $v_3$ the triangular flow.

\begin{figure}[t]
\centering
\raisebox{6mm}{\subfloat{\includegraphics[width=0.58\linewidth]{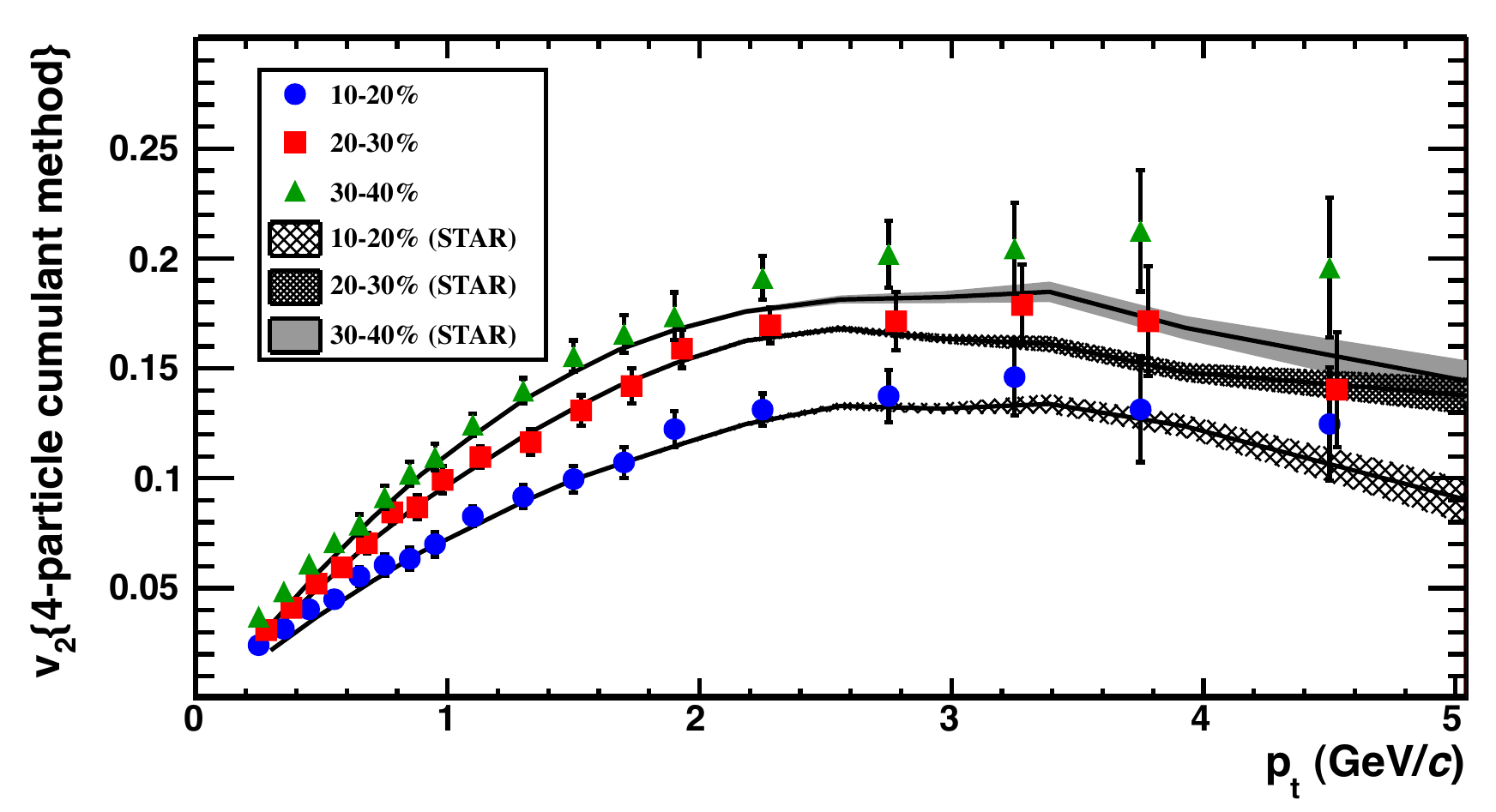}}}
\subfloat{\includegraphics[width=0.41\linewidth]{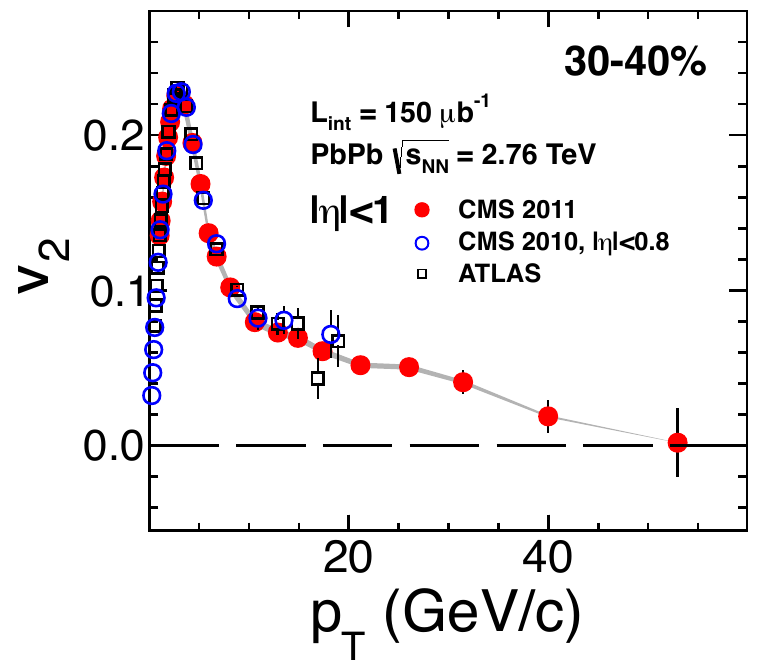}}
\caption{a) Elliptic flow coefficient $v_2$ for charged particles as a function of $p_T$ for three centrality classes measured by ALICE \cite{Aamodt:2010pa}. For the shown classes $v_2$ decreases with increasing centrality. The $p_T$ and centrality dependence at LHC and RHIC is remarkably similar. b) $v_2$ for charged particles up to $p_T \approx \unit[50]{GeV}/c$ from CMS \cite{Chatrchyan:2012xq}.}
\label{fig:v2}
\end{figure}
Despite the large difference in $\sqrt{s_{NN}}$ the $p_T$ and centrality dependence of the elliptic flow at RHIC and LHC are very similar, as shown in Fig.~\ref{fig:v2}a. The $v_2$ coefficient was determined with the 4-particle cumulant method in order to minimize the contribution of non-flow effects, e.g., mini-jet production. Owing to the larger mean transverse momentum of the charged particles at the LHC, the $p_T$-integrated $v_2$ at the LHC is about 30\,\% larger than at RHIC.

The hydro picture for $v_2$ is only valid at low transverse momentum ($p_T \lesssim \unit[1.5]{GeV}/c$). However, a non-vanishing $v_2$ was measured for much larger $p_T$, see Fig.~\ref{fig:v2}b. At large $p_T$ the $v_2$ is believed to result from the path length dependence of the energy loss of quark and gluon jets in the created medium (cf. Sec.~\ref{sec:jet_quenching}). The path length in the direction of the reaction plane is shorter than perpendicular to it, resulting in $v_2 > 0$.

\begin{figure}[t]
\centering
\subfloat{\includegraphics[width=0.465\linewidth]{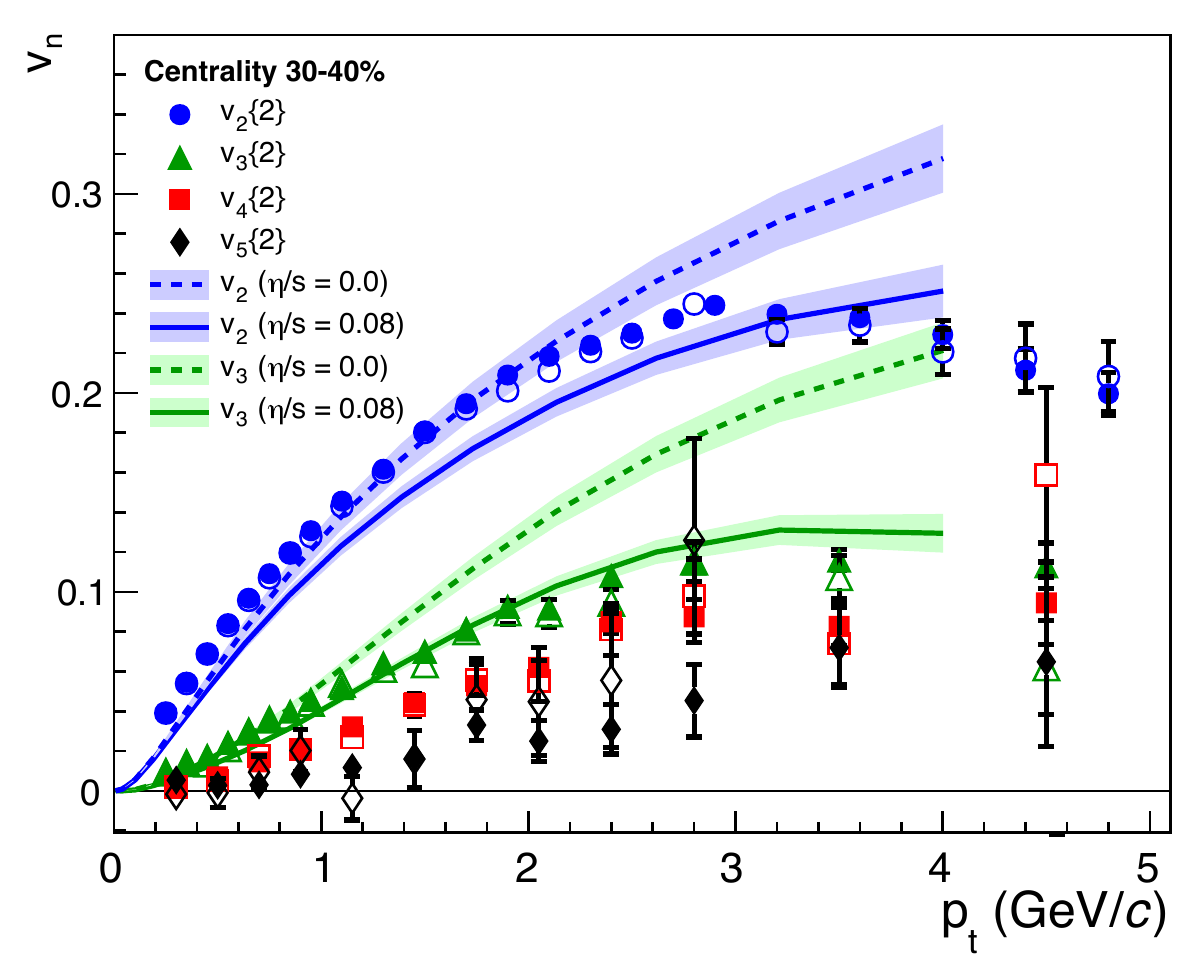}}
\subfloat{\includegraphics[width=0.535\linewidth]{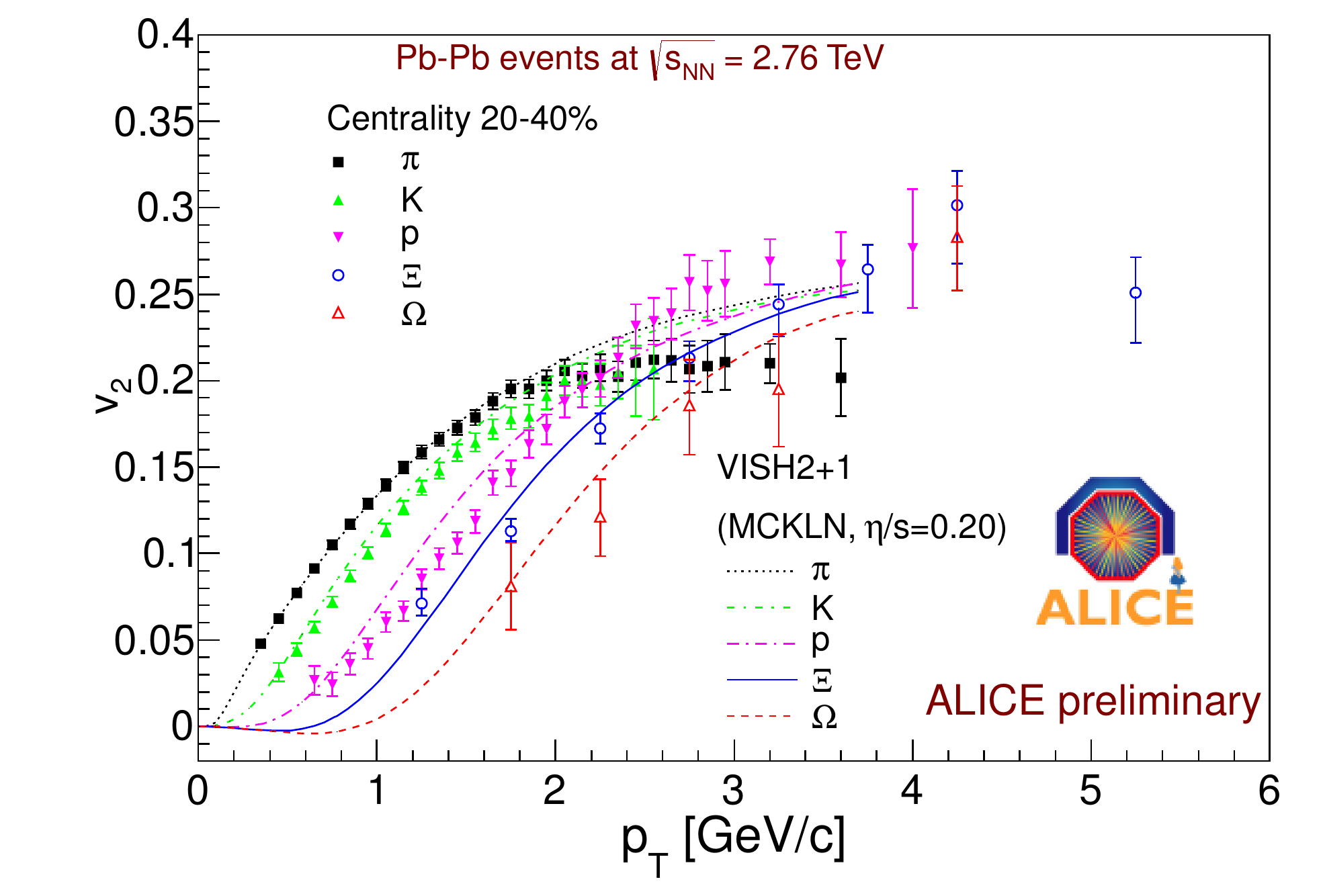}}
\caption{Fourier coefficients $v_2$, $v_3$, $v_4$, and $v_5$  of charged particles as a function of $p_T$ in Pb+Pb collisions ($30-40$\,\% most central class) from ALICE \cite{ALICE:2011ab}. The $p_T$ dependence of $v_2$ and $v_3$ is described with hydro calculations with small $\eta/s$. b) $v_2$ as a function of $p_T$ for different particles species from ALICE. The $v_2$ shows the mass ordering expected from hydro calculations.}
\label{fig:v3v4v5_and_idv2}
\end{figure}
The measurement of the Fourier coefficients, $v_n$, provides limits on the shear viscosity to entropy density ratio $\eta/s$ of the created medium. In addition to $v_2$, flow coefficient up to $v_5$ have been measured (Fig.~\ref{fig:v3v4v5_and_idv2}a). The effect of viscosity is to dissipate initial pressure gradients and to reduce collective flow. This can be seen from the hydro calculations in Fig.~\ref{fig:v3v4v5_and_idv2}a for different $\eta/s$ values. By measuring higher flow coefficients like $v_3$ in addition to $v_2$ the sensitivity for $\eta/s$ is increased. Fig.~\ref{fig:v3v4v5_and_idv2}a shows that for certain assumption about the initial distribution of energy density (``Glauber initial conditions'') the data at low $p_T$ are described with $\eta/s \approx 1/4\pi$.
Based on LHC data the current upper bound is $\eta/s \lesssim 2/(4\pi) =2 \times (\eta/s)_\mathrm{min}$ \cite{Muller:2012zq}. 

One of the most compelling pieces of evidence for the hydrodynamic expansion of the created medium comes from the observed mass ordering of the $v_2$ of identified particles. The $p_T$ dependence of elliptic flow expected from hydro can be approximated as $v_2 \sim (p_T-\beta m_T)/T$ where $\beta$ is the average transverse flow velocity and $m_T = \sqrt{p_T^2+m^2}$ the transverse mass. The mass ordering is indeed observed and described by hydro calculations as shown in Fig.~\ref{fig:v3v4v5_and_idv2}b.

\section{Jet Quenching}
\label{sec:jet_quenching}
Heavy-ion physics at LHC energies benefits from the abundant production of hard probes. Hard probes are useful because they are produced in the early stage of the collisions, prior to the formation of the QGP. Moreover, their initial production rate is calculable with perturbative QCD, making them a ``calibrated'' probe. Jet quenching, i.e., the energy loss of quarks and gluons from hard scattering processes, was discovered at RHIC by measuring single particle yields at high $p_T$. The interest in observables related to jet quenching is two-fold: one would like to 1) understand the mechanism of parton energy loss and 2) use hard probes as a tool to characterize the QGP.

\begin{figure}
\centering
\subfloat{\includegraphics[width=0.5\linewidth]{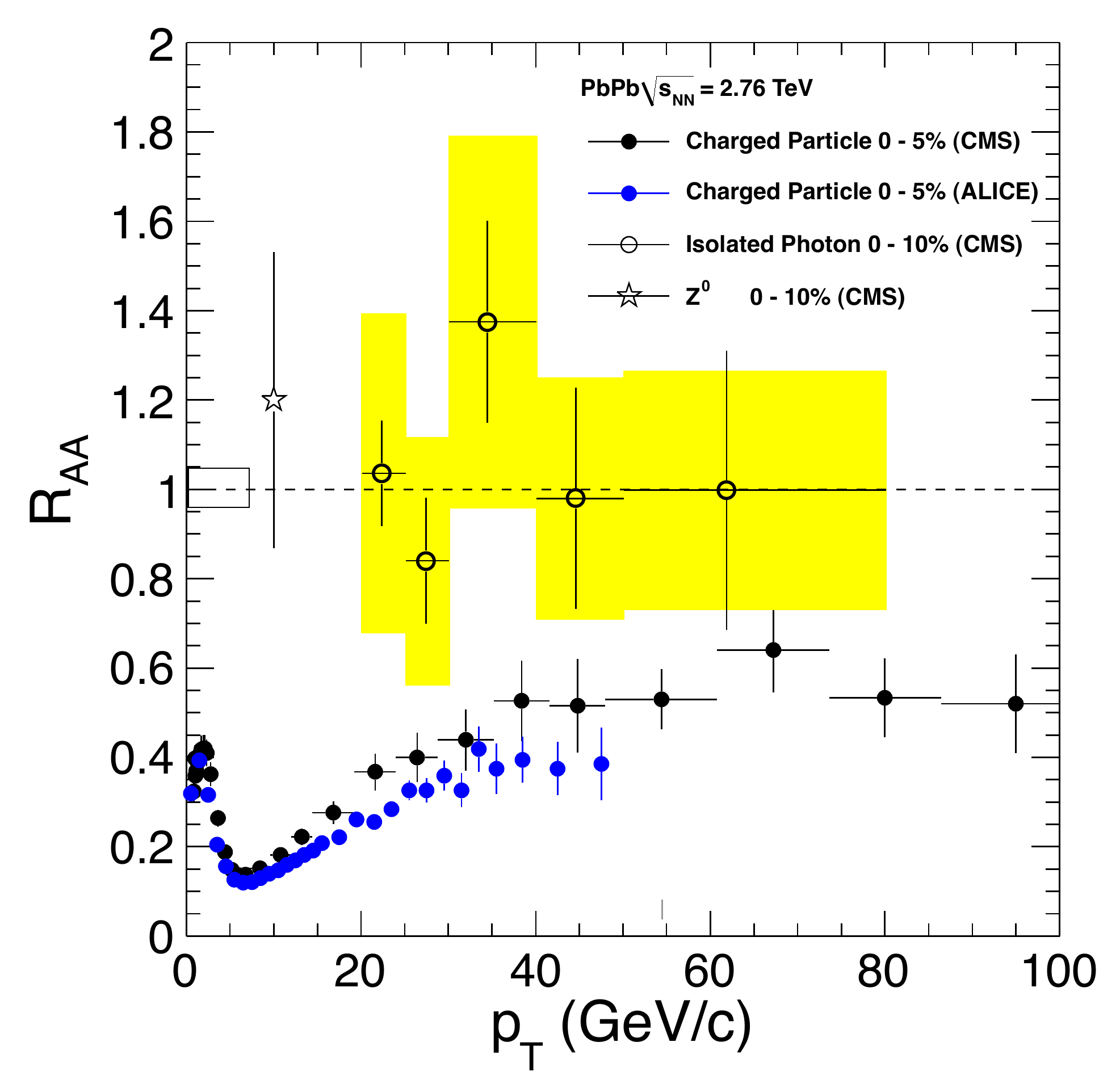}}
\hfill
\raisebox{10mm}{\subfloat{\includegraphics[width=0.45\linewidth]{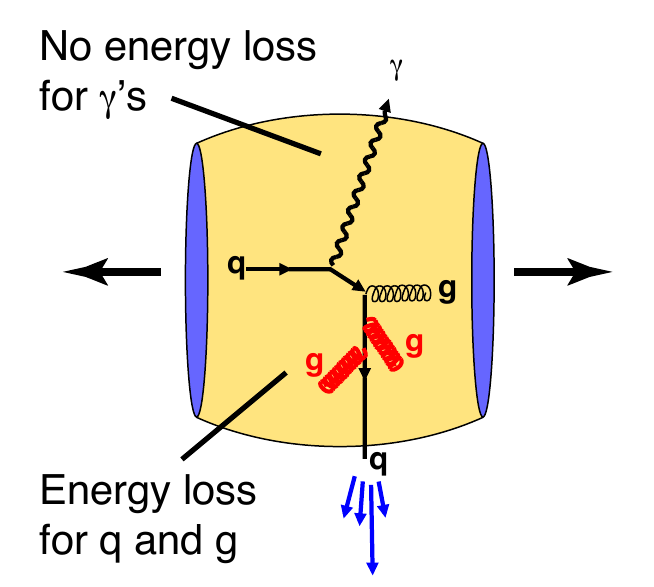}}}
\caption{$R_{AA}(p_T)$ for charged particles and isolated photons in central Pb+Pb collisions at the LHC \cite{Muller:2012zq}. Charged particles are suppressed whereas isolated photon are not. This is expected in the parton energy loss picture because photons essentially do not interact with the medium as indicated by the sketch.}
\label{fig:raa}
\end{figure}

Charged particles yields in central Pb+Pb collisions at $\sqrt{s_{NN}} = \unit[2.76]{TeV}$ are suppressed by more than a factor of 5 at $p_T \approx \unit[7]{GeV}/c$, see Fig.~\ref{fig:raa}. The suppression is quantified with the nuclear modification factor
\begin{equation}
R_{AA}=\frac{\mathrm{d}N/\mathrm{d}p_T (A+A)}
{\langle T_{AA} \rangle \times \mathrm{d}\sigma/\mathrm{d}p_T (p+p)}
\end{equation}
where the nuclear overlap function $\langle T_{AA} \rangle = \langle N_\mathrm{coll} \rangle / \sigma_\mathrm{inel}^{pp}$ describes the increase of the parton flux from p+p to A+A. Without nuclear effects $R_{AA} = 1$ in the hard scattering regime ($p_T \gtrsim \unit[2]{GeV}/c$).  $R_{AA}$ depends on the parton energy loss as well as on the steepness of the parton spectrum. Thus, the same $R_{AA}$ at different $\sqrt{s_{NN}}$ corresponds to a different energy loss.

The rise of $R_{AA}$ with $p_T$ was for the first time firmly established at the LHC. The large $p_T$ reach of the LHC data helps unveil the dependence of the energy loss on the initial parton energy. The data are consistent with a decrease of the fractional energy loss $\Delta E/E$ with increasing parton energy $E$ as expected in energy loss models based on perturbative QCD. The nuclear modification factor may also be affected by initial state effects like gluon shadowing. These effects will be studied in the p+Pb run in fall of 2012. 

Prompt photons provide a crucial test for parton energy loss model as they do not interact with the medium via the strong interaction. Therefore, they are expected to leave the fireball unscathed. The same holds for $Z$ bosons. Fig.~\ref{fig:raa} shows that isolated photons and also Z bosons are indeed not suppressed ($R_{AA} \approx 1$). This provides further evidence for the parton energy loss picture.

\begin{figure}[t]
\centering
\subfloat{\includegraphics[width=0.48\linewidth]{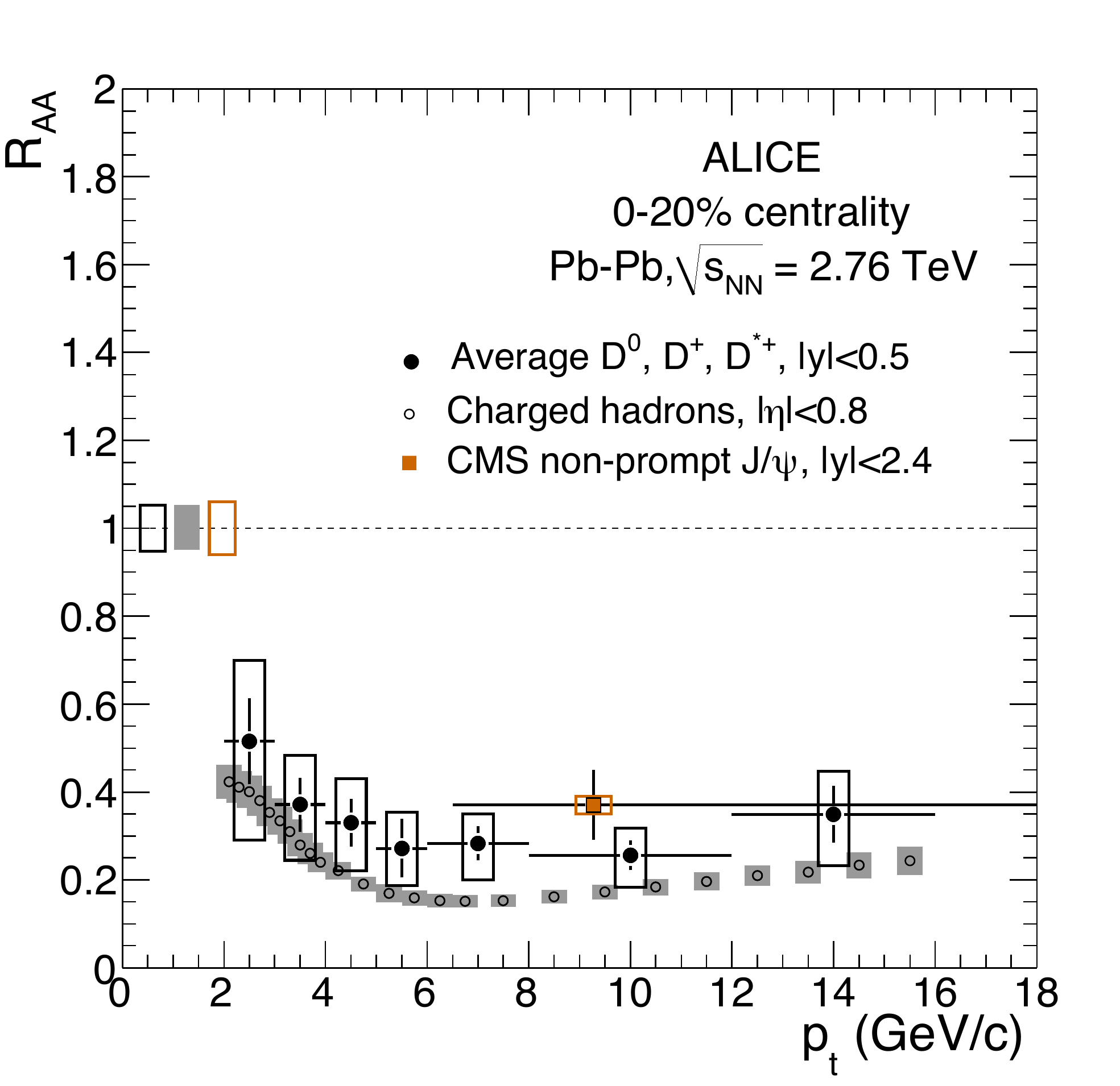}}
\hfill
\subfloat{\includegraphics[width=0.48\linewidth]{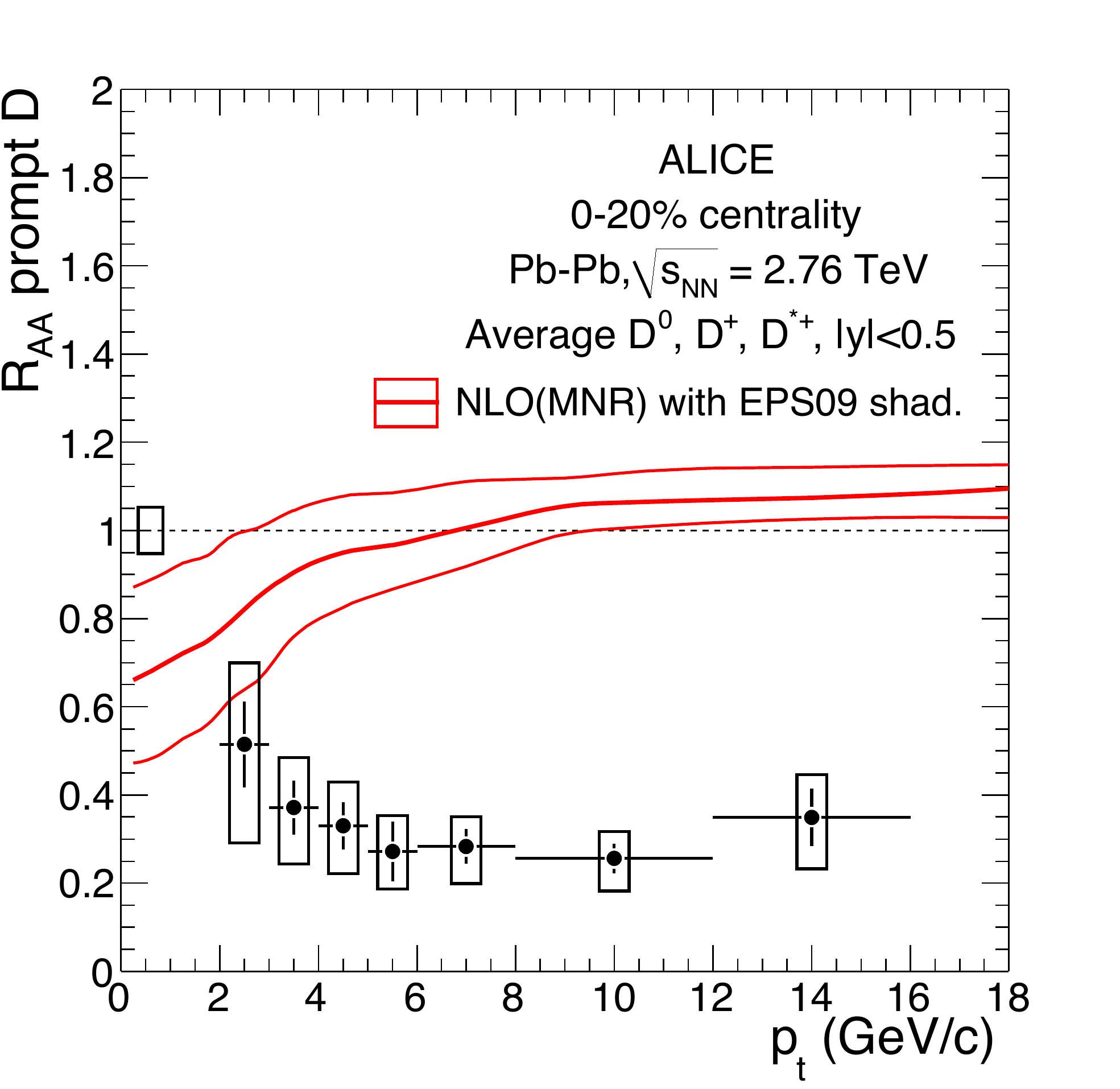}}
\caption{ a) The comparison of the $R_{AA}$ for charged hadron, D mesons, and $J/\psi$'s from B meson decays (``non-prompt $J/\psi$'s'') tests the expected hierarchy $\Delta E_g > \Delta E_{u,d,s} > \Delta E_c > \Delta E_b$. b) Initial state effects related to modification of the gluon distribution function in the Pb nucleus (shadowing) appear to be negligible for $p_T \gtrsim \unit[5]{GeV}/c$ as indicated by the NLO perturbative QCD calculation.}
\label{fig:d_meson_raa}
\end{figure}
Radiative energy loss, i.e., gluon emission induced by the medium, is expected to be the dominant energy loss mechanism in the QGP. In this picture the energy loss of gluons, light and heavy quarks in a QGP is expected to exhibit the ordering $\Delta E_g > \Delta E_{u,d,s} > \Delta E_c > \Delta E_b$ corresponding to $R_{AA}(\pi) < R_{AA}(\mbox{D mesons}) < R_{AA}(\mbox{B mesons})$. Note that for $p_T \lesssim \unit[50]{GeV}/c$ pions predominantly originate from the fragmentation of gluons jets.
The larger energy loss for gluons relative to quarks is due to the different color factor, $C_F = 3$ for gluons and $C_F = 4/3$ for quarks. The smaller energy loss of heavy quarks with respect to light quarks is due to the dead-cone effect \cite{Dokshitzer:2001zm}.

The suppression for charged hadrons, prompt D mesons, and B mesons in central Pb+Pb collisions at $\sqrt{s_{NN}} = \unit[2.76]{TeV}$ is found to be rather similar as shown in Fig.~\ref{fig:d_meson_raa}a. Even though the differences are small, there is at least an indication for the expected ordering. The $R_{AA}$ of B mesons in  Fig.~\ref{fig:d_meson_raa}a was determined by measuring the $R_{AA}$ of non-prompt $J/\psi$'s. The effect of shadowing is estimated in Fig.~\ref{fig:d_meson_raa}b with a NLO perturbative QCD calculation which employs the EPS09 parton distribution. Gluon shadowing appears to contribute to the suppression of D mesons only for $p_T \lesssim \unit[5]{GeV}/c$.

\begin{figure}
\centering
\subfloat{\includegraphics[width=0.48\linewidth]{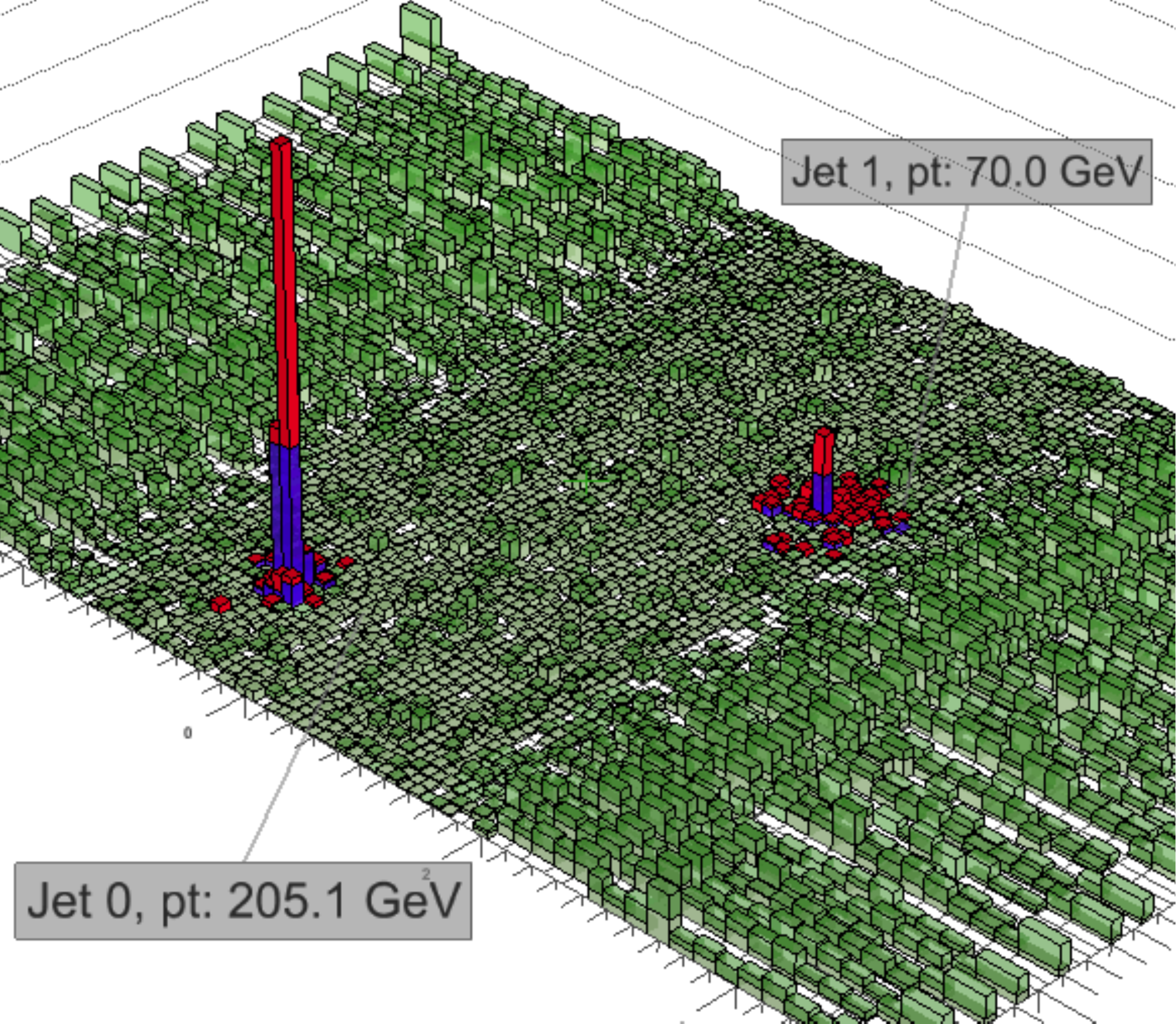}}
\subfloat{\includegraphics[width=0.52\linewidth]{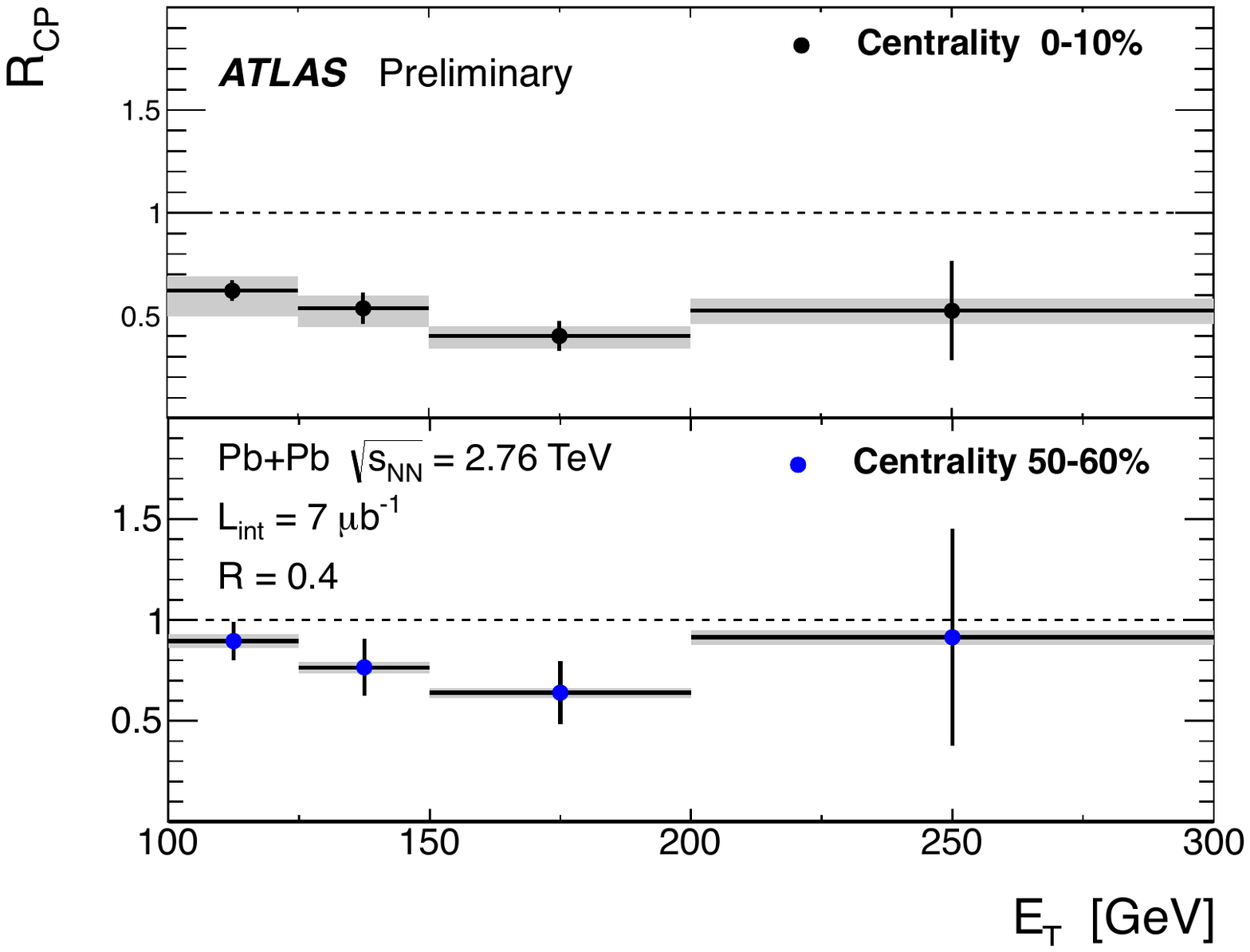}}
\caption{a) Calorimeter energy vs. $\eta$ and $\varphi$ in a Pb+Pb collision at $\sqrt{s_{NN}} = \unit[2.76]{TeV}$ containing a di-jet with large energy asymmetry measured by CMS \cite{Chatrchyan:2011sx}. b) $R_{CP}$ (see text) of jet spectra determined with an anti-$k_T$ algorithm with a radius parameter $R=0.4$. For the $0-10$\,\% most central Pb+Pb events the jet spectrum is suppressed indicating that with $R=0.4$ the energy of jets suffering energy loss is not fully recovered.}
\label{fig:jets}
\end{figure}
The large cross section for hard processes at the LHC provide a unique opportunity to study parton energy loss with fully reconstructed jets \cite{Aad:2010bu,Chatrchyan:2011sx}. ATLAS and CMS studied di-jet production in Pb+Pb collisions and find energy differences between the leading and sub-leading jet much larger than in pp collisions. This can be naturally explained with di-jet produced close to the edge of the overlap zone so that one parton has a long path length in the medium and loses energy whereas the other escapes without energy loss. A closer look at di-jet production in Pb+Pb collisions reveals that the di-jets are still produced back-to-back (i.e., there is no angular decorrelation) and that the momentum distribution of jet particles transverse to the jet axis is like in pp collisions. So one has to ask: Where does the energy of jets with reduced energy go?

A first step to address this question is to study the modification of jet spectra in central Pb+Pb collisions relative to peripheral collisions with the aid of ratio $R_{CP} = N_\mathrm{coll}^\mathrm{60-80\%}/N_\mathrm{coll}^\mathrm{cent} \times   (dN_\mathrm{jet}^\mathrm{cent}/dE_T)/(dN_\mathrm{jet}^\mathrm{60-80\%}/dE_T)$ where $dN_\mathrm{jet}/dE_T$ is the jet $E_T$ spectrum normalized per event. ATLAS finds a factor of 2 suppression ($R_{CP} \approx 0.5$) in central Pb+Pb collisions using an anti-$k_T$ algorithm with a radius parameter of $R=0.4$ \cite{Cole:2011zz}. This indicates that with this radius parameter the full jet energy is not recovered. A detailed study by CMS in which particle tracks are correlated with the axis of the leading jet shows that the energy difference in di-jets is balanced by low $p_T$ particles ($0.5 \lesssim p_T < \unit[2]{GeV}/c$) at large angles relative to the axis of the sub-leading jet \cite{Chatrchyan:2011sx}.

\section{Quarkonia}
\begin{figure}
\centering
\raisebox{1mm}{\subfloat{\includegraphics[width=0.56\linewidth]{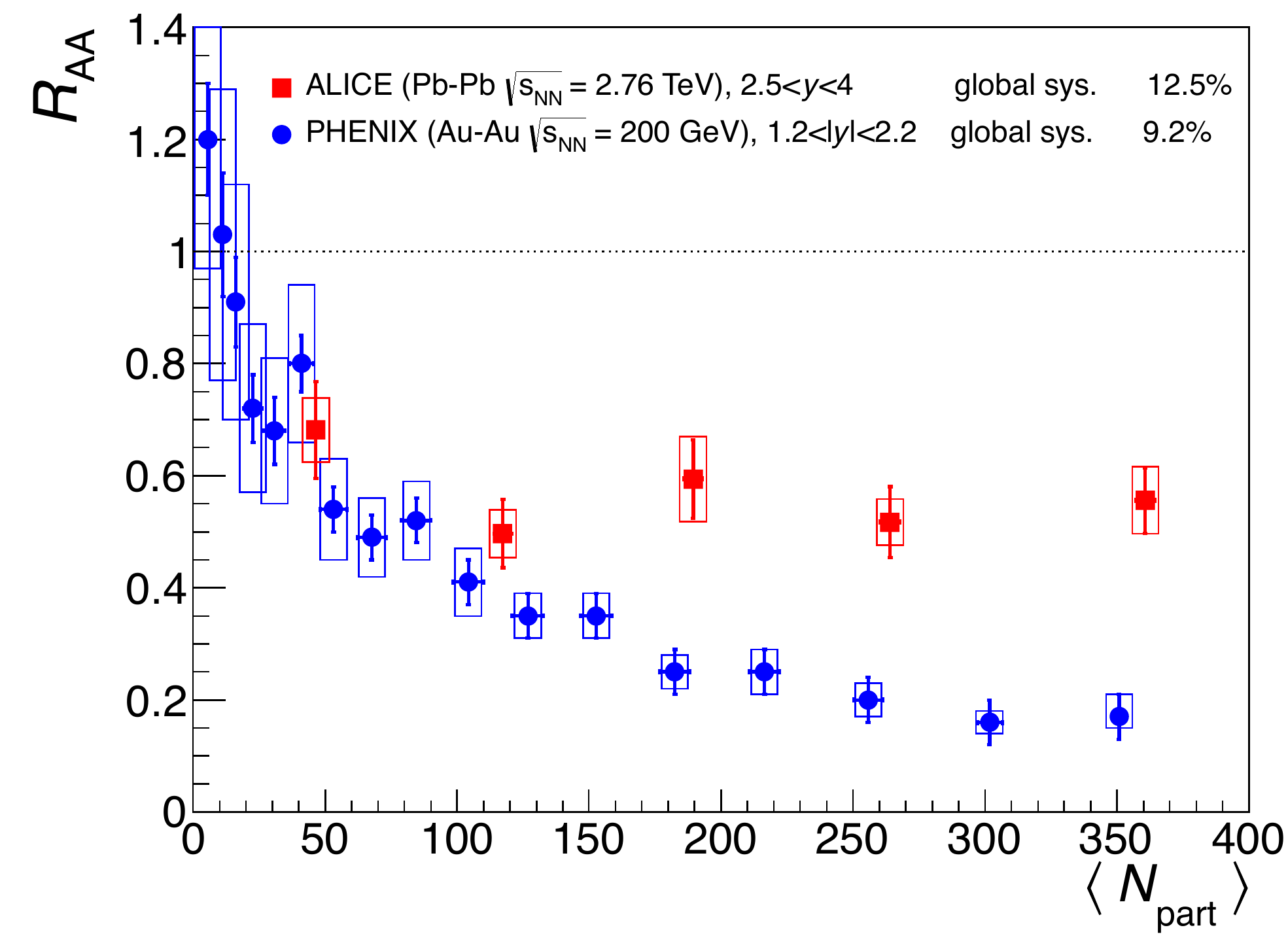}}}
\subfloat{\includegraphics[width=0.43\linewidth]{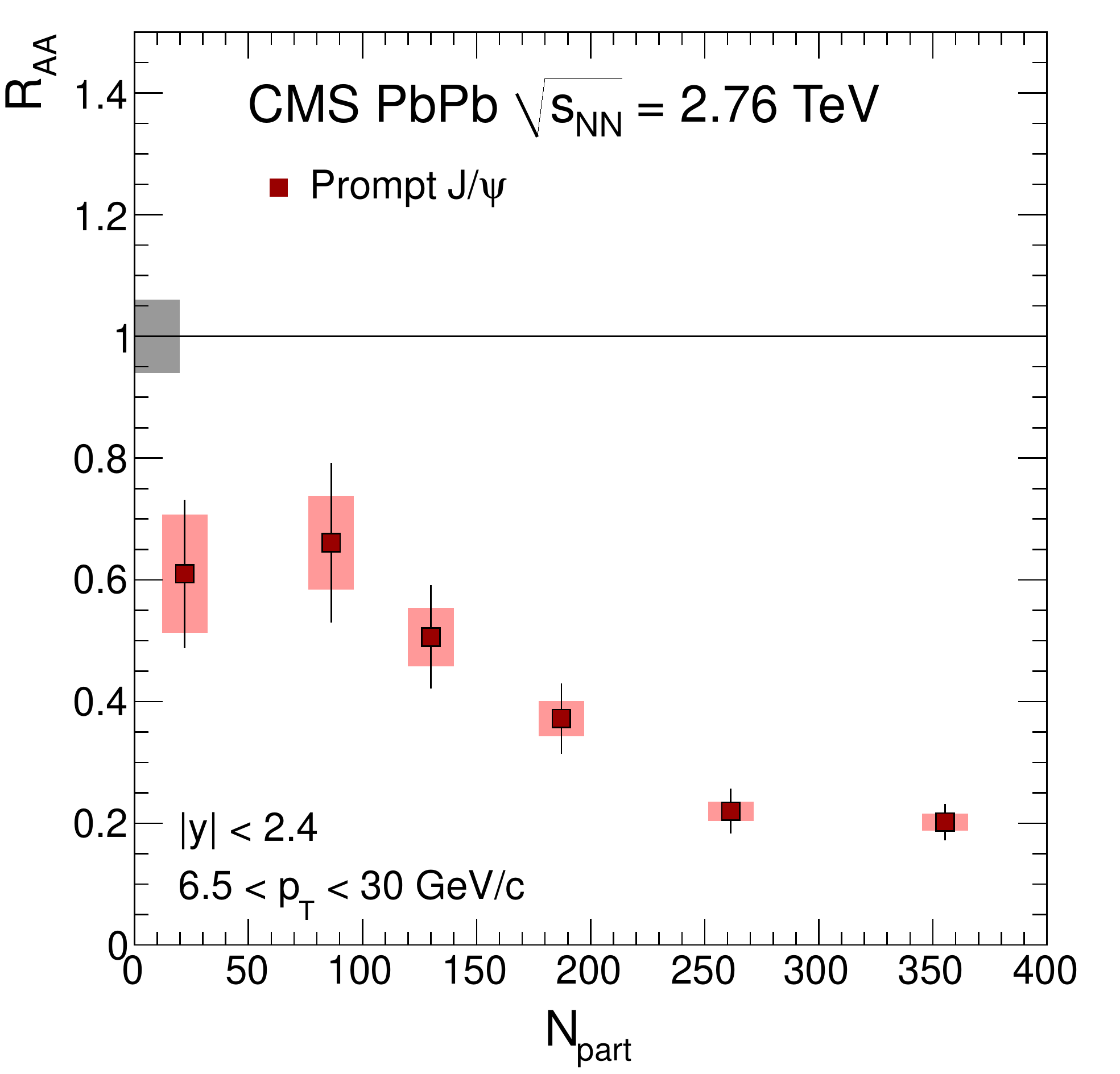}}
\caption{a) $R_{AA}$ of inclusive $J/\psi$'s measured by ALICE over the full $p_T$ range ($p_T>0$) and in $2.5 < y < 4$ as a function of centrality \cite{Abelev:2012rv}. In central collisions the suppression ($R_{AA} \approx 0.5-0.6$) is smaller than at RHIC. b) At larger $p_T$ ($p_T>\unit[6.5]{GeV}/c$) prompt $J/\psi$ measured by CMS exhibit a stronger suppression in central collisions ($R_{AA} \approx 0.2$) \cite{Chatrchyan:2012np}.}
\label{fig:quarkonia}
\end{figure}
Quarkonia belong to the classical QGP probes \cite{Matsui:1986dk}. Considering the total $J/\psi$ yield, the formation of a QGP in A+A collisions at low $\sqrt{s_{NN}}$ (e.g. at CERN SPS energies) is expected to result in a suppression whereas at higher $\sqrt{s_{NN}}$ the formation of a QGP may lead to a less strong suppression or even an enhancement \cite{BraunMunzinger:2007zz}. Color screening  is expected to prevent the binding of $c \bar c$ and $b \bar b$ pairs in deconfined matter. The dissociation temperature $T_D$ is different for different quarkonium states, e.g., $T_D(J/\psi) \approx 1.2\,T_c$, $T_D(\psi') \approx T_c$, and $T_D(\Upsilon) \approx 2 \,T_c$ \cite{Chatrchyan:2012np}. Therefore, information about the temperature of the QGP can be obtained from the comparison of yields of different quarkonium states. At the LHC, on the order of 100 $c \bar c$ pairs are produced in a central Pb+Pb collision. It it thus conceivable that $J/\psi$'s in A+A collisions are produced at the phase transition due to statistical recombination of $c \bar c$ pairs.

For the total inclusive yields of $J/\psi$'s (i.e., integrated over the full range $p_T >0$) ALICE measured $R_{AA} \approx 0.6$ independent of the centrality of the Pb+Pb collision (Fig.~\ref{fig:quarkonia}a).  Note that the contribution of non-prompt $J/\psi$'s from B meson feed-down to the inclusive $J/\psi$ yield is $\sim 15$\,\%. Gluon shadowing alone is expected to lead to a $J/\psi$ suppression of $R_{AA} \approx 0.8$. This indicates that there is only a moderate final state suppression. Interestingly, the suppression in central Pb+Pb collisions is smaller than at RHIC. For prompt $J/\psi$'s at larger $p_T$ ($> \unit[6.5]{GeV}/c$) the suppression increases with centrality up to $R_{AA} \approx 0.2$ in central collisions. All in all, these observations are in qualitative agreement with the recombination picture.

\section{Conclusions}
Owing to the large initial energy density, the long QGP lifetime, and the abundant production of hard probes the LHC is ideal for studying the QGP. Particles produced in Pb+Pb collisions at the LHC exhibit strong radial and anisotropic flow confirming the standard reaction scenario. The medium created at the LHC has the same ``perfect liquid'' properties as found at RHIC. The medium is opaque to jets and the large $p_T$ reach of the $R_{AA}$ measurements at the LHC provides constraints for parton energy loss models. The data on charmonium production are consistent with a statistical formation of $J/\psi$ at the phase transition.



{\raggedright
\begin{footnotesize}
\bibliographystyle{DISproc}
\bibliography{reygers_klaus.bib}
\end{footnotesize}
}


\end{document}